\documentclass[12pt,twoside, draftclsnofoot, onecolumn, journal]{IEEEtran}
\IEEEoverridecommandlockouts
\usepackage{amsfonts}
\usepackage{amsmath}
\usepackage{theorem}
\usepackage{graphicx}
\usepackage{epsfig}
\usepackage{cite}
\usepackage{epstopdf}
\usepackage{algorithm}
\usepackage{algorithmic}
\usepackage{multicol}
\usepackage{amssymb}

\newcommand{\defeq}{\mathrel{\mathop:}=}

\newtheorem{theorem}{Theorem}
\newtheorem{lemma}{Lemma}
\newtheorem{proposition}{Proposition}

\newtheorem{assumption}{Assumption}
\newtheorem{remark}{Remark}
\newcommand{\boxedeqn}[1]{%
    \[\fbox{%
        \addtolength{\linewidth}{-2\fboxsep}%
        \addtolength{\linewidth}{-2\fboxrule}%
        \begin{minipage}{\linewidth}%
        \begin{small}%
        \begin{align}#1\end{align}%
        \end{small}%
        \end{minipage}%
      }\nonumber\]%
  }
\DeclareMathOperator*{\argmin}{arg\,min}


\begin{document}

\title{Distributed Output-Feedback LQG Control with Delayed Information Sharing}
\author{Hamid Reza Feyzmahdavian, Ather Gattami, and Mikael Johansson
\thanks{H. R. Feyzmahdavian, A. Gattami, and M. Johansson are with ACCESS Linnaeus Center, School of Electrical
Engineering, KTH-Royal Institute of Technology, SE-100 44 Stockholm, Sweden.
E-mails: {\{hamidrez, gattami, mikaelj\}@kth.se}}
}

\maketitle


\begin{abstract}
This paper develops a controller synthesis method for distributed LQG control problems under output-feedback. We consider a system consisting of three interconnected linear subsystems with a delayed information sharing structure. While the state-feedback case  has previously been solved, the extension to output-feedback is nontrivial as the classical separation principle fails. To find the optimal solution, the controller is decomposed into two independent components: a centralized LQG-optimal controller under delayed state observations, and a sum of correction terms based on additional local information available to decision makers. Explicit discrete-time equations are derived whose solutions are the gains of the optimal controller.\footnote{A preliminary version of this work was presented in~\cite{Feyzmahdavian:12-1}}
\end{abstract}


\section{Introduction}

Control with information constraints imposed on decision makers, sometimes called team theory or distributed control, has been very challenging for decision theory researchers.  In general, several classes of these problems are currently computationally intractable \cite{blondel:2000}. Early work \cite{witsenhausen:1968} showed that even in a simple static linear quadratic decision problem, complex nonlinear decisions could outperform any given linear decision. As a result, much research has focused on identifying classes of decentralized control problems that are tractable \cite{ho:chu,bamieh:02,shah:10,Feyzmahdavian:12}.

Distributed Linear Quadratic Gaussian (LQG) control with communication delays has a rich literature dating back to the $\mbox{1970s}$. Even though the LQG problem under one-step  delay information sharing pattern has been solved in~\cite {Sandell:74,Kurtaran:74,Toda:75,Yoshi:75}, generalizing their approaches to other delay structures is non-trivial. In \cite{rantzer:cdc06} and \cite{gattami:tac:10}, a computationally efficient solution for the LQG output-feedback problem with communication delays was presented using a state space formulation and covariance constraints, but the controller structure is not apparent from the corresponding semi-definite programming solution. In \cite{lamperski:cdc11}, the authors consider LQG control with communication delays for the three interconnected systems. While they provide an explicit solution, their approach is restricted to state-feedback and assumes independence of disturbances acting on each subsystem.

In this paper, we generalize the results in \cite{lamperski:cdc11} to output-feedback and correlated disturbances. We consider three interconnected systems over a strongly connected graph, which implies information from neighbors is available with one step delay and the global information is available to all decision makers with two step delay. We derive an output-feedback law that minimizes a finite-horizon quadratic cost. The problem considered here provides the fundamental understanding for general delay structures.

The main contribution of this paper is the explicit state-space realization of the LQG output-feedback problems with communication delays.
The problem is solved by decomposing the controller into two components. One is the same as centralized LQG problem under two-step information delay and the other is the sum of correction terms based on local information available to decision makers. Specifically, the optimal control has the form
\begin{align*}
u(k) = F(k)\bigl(y(k)-C\widehat{x}^{[1]}(k)\bigr)+F^{[1]}(k)\bigl(y(k-1)-C\hat{x}(k-1\vert k-2) \bigr)+ L(k)\hat{x}(k\vert k-2),
\end{align*}
where $\hat{x}(k-1\vert k-2)$ and $\hat{x}(k\vert k-2)$ is the one- and two-step estimation of the state based on the common two-step delayed information, and $\widehat{x}^{[1]}(k)$ is an improved state estimate based on local information up to time~$k-1$ available to decision makers at time $k$. While the gain matrix $L$ might be full (in fact, it is the standard LQR gain computed via discrete-time Riccati recursion), the gain matrices $F$ and $F^{[1]}$ have a sparsity structure that complies with the information constraints. We further show that $F$ and $F^{[1]}$ can be computed via convex programming.

The paper is organized as follows. Section~\ref{sec:Problem} defines the general problem studied in this paper.
In Section~\ref{sec:Estimation}, we review the standard discrete time Kalman filter and derive an optimal estimation algorithm for the three-player problem.
In Section~\ref{sec:Decomposition}, it is shown that the three-player control problem can be separated into two optimization problems.
The main result of this paper is stated in Section~\ref{sec:Decomposition1}. Numerical results are given in Section~\ref{sec:Example} and finally conclusions and future work are outlined in Section~\ref{sec:Conclusion}.


\subsection{Notation}
\label{sec:preliminaries}

Throughout the paper, we use the following notation: matrices are written in uppercase letters and vectors in lowercase letters. The sequence $x(0)$, $x(1)$, $\ldots$ , $x(k)$ is denoted by $x(0:k)$. The symbol $I$ denotes the identity matrix whose size can be determined from its context. For a matrix $X$ partitioned into blocks, $[X]_{S_1S_2}$ denotes the sub-matrix of $X$ containing exactly those rows and columns corresponding to the sets $S_1$ and $S_2$, respectively. For instance $[X]_{\{1\}\{2,3\}}=\begin{bmatrix} X_{12} & X_{13}\end{bmatrix}$. The trace of a square matrix $X$ is denoted by $\textbf{Tr}\{X\}$. Given $A\in\mathbb{R}^{m\times n}$, we can write $A$ in terms of its columns as
$A=\begin{bmatrix} a_1 & \cdots & a_n\end{bmatrix}$. Then operation $\textup{vec}(A)$ results in an $mn \times 1$ column vector
\begin{align*}
\textup{vec}(A)=\begin{bmatrix} a_1 \\ \vdots \\ a_n\end{bmatrix}.
\end{align*} For $A\in\mathbb{R}^{m\times n}$ and $B\in\mathbb{R}^{r\times s}$, the operation $A\otimes B \in\mathbb{R}^{mr\times ns}$ denotes the \textit{Kronecker product} of $A$ and $B$. We denote the expectation of a random variable $x$ by $\textbf{E}\{x\}$. The conditional expectation of $x$ given $y$ is denoted by $\textbf{E}\{x|y\}$. The covariance of zero-mean random vectors $x$ and $y$, defined by $\textbf{E}\{xy^T\}$, is denoted by $\textbf{{Cov}}\{x,y\}$.


\section{Problem Formulation}
\label{sec:Problem}

Consider the following linear discrete time system composed of $m$ interconnected subsystems
\begin{align}
\begin{split}
x_i(k+1)&=\sum_{j=1}^m A_{ij}x_j(k)+B_iu_i(k)+w_i(k)\\
y_i(k)&=C_ix_i(k)+v_i(k),
\end{split}
\label{mainsystem3}
\end{align}
for $i=1,\ldots,m$. Here, $x_{i}\in\mathbb{R}^{n_i}$ is the state , $u_{i}\in \mathbb{R}^{q_i}$ is the control signal, $y_{i}\in \mathbb{R}^{p_i}$ is the measurement output, $w_i$ is the disturbance, and $v_i$ is the measurement noise of subsystem $i$. Here, $A_{ij}\in\mathbb{R}^{n_i\times n_j}$, $B_{i}\in\mathbb{R}^{n_i\times q_i}$ and $C_{i}\in\mathbb{R}^{p_i\times n_i}$ are constant matrices. Let us define
\begin{align*}
x=\begin{bmatrix} x_{1} \\  \vdots \\  x_{m}\\ \end{bmatrix},\;u=\begin{bmatrix} u_{1} \\  \vdots \\  u_{m}\\ \end{bmatrix},\;
y=\begin{bmatrix} y_{1} \\  \vdots \\  y_{m}\\ \end{bmatrix},\;
w=\begin{bmatrix} w_{1} \\  \vdots \\  w_{m}\\ \end{bmatrix},\;v=\begin{bmatrix} v_{1} \\ \vdots \\  v_{m}\\ \end{bmatrix}.
\end{align*}
Then the system dynamics (\ref{mainsystem3}) can be written as
\begin{align}
\begin{split}
x(k+1)&= Ax(k)+ B u(k)+w(k)\\
y(k)&= C x(k)+v(k),
\end{split}
\label{augmentedsystem}
\end{align}
where $A=[A_{ij}]\in\mathbb{R}^{n\times n}$, $B=\textbf{diag}(B_1,\ldots,B_m)\in\mathbb{R}^{n\times q}$ and $C=\textbf{diag}(C_1,\ldots,C_m)\in\mathbb{R}^{p\times n}$. Both $w$ and $v$ are assumed to be Gaussian white noises with covariance matrix
\begin{eqnarray*}
\textbf{E} \left\{\begin{bmatrix} {w(k)} \\ {v(k)} \end{bmatrix}{\begin{bmatrix} {w(l)} \\ {v(l)} \end{bmatrix}}^{T}\right\}=\delta(k-l)\begin{bmatrix}  W & 0 \\ 0 &  V \end{bmatrix},
\end{eqnarray*}
where $\delta(k-l)=1$ if $k=l$ and $\delta(k-l)=0$ if $k\neq l$.
\begin{assumption}
$V$ is positive definite.
\label{noisematrix}
\end{assumption}

The interconnection structure of system (\ref{augmentedsystem}) can be represented by a graph $\mathcal{G}$ whose nodes correspond to subsystems. The graph $\mathcal{G}$ has an arrow from node $j$ to node $i$ if and only if $ A_{ij} \neq 0$ (\emph{i.e.} if $x_{j}(k)$ influences $x_i(k+1)$). Assume that $\mathcal{G}$ is strongly connected and passing information from one node to another along the graph takes one time step. Let $d_{ij}$ be the length of the shortest path from node $i$ to node $j$ with $d_{ii}=0$. Then node $i$ receives the information available to node $j$ after $d_{ji}$ time steps, and hence the available information set of subsystem $i$ at time $k$ is given by
\begin{eqnarray}
\mathcal{I}_{i}(k)=\bigl\{y_1(0:k-d_{1i}),\;\ldots\;,y_i(0:k),\;\ldots\;,y_m(0:k-d_{mi})\bigr\}.
\label{historysetc}
\end{eqnarray}
The control problem is to minimize finite-horizon cost
\begin{eqnarray}
J=\textbf{E}\left\{\sum_{k=0}^{N-1}{\begin{bmatrix} {x(k)} \\ {u(k)}\end{bmatrix}}^{T}Q{\begin{bmatrix} {x(k)} \\ {u(k)}\end{bmatrix}}+x(N)^{T}Q_{0}x(N)\right\},
\label{cost}
\end{eqnarray}
subject to inputs of the form
\begin{align*}
u_i(k)=\mu_i\bigl(\mathcal{I}_i(k)\bigr),\; i=1,\ldots,m,
\end{align*}
where $\mu_i$ is the Borel-measurable function. Matrix $Q$ is partitioned according to the dimensions of $x$ and $u$ as
$$
Q=\begin{bmatrix} Q_{xx} &  Q_{xu}\\ {Q}^{T}_{xu} &  Q_{uu}\end{bmatrix}.
$$
\begin{assumption}
The matrices $Q_0$ and $Q$ are positive semi-definite, and $Q_{uu}$ is positive definite.
\label{costmatrix}
\end{assumption}

The information structure (\ref{historysetc}) can be viewed as the consequence of delays in the communication channels between the
controllers. The assumptions about the information structure and the sparsity of dynamics guarantee that information propagates at least as fast as the dynamics on the graph. This information pattern is a  simple case of \textit{partially nested} information structure that has been studied in \cite {ho:chu}. The optimal controller with this information pattern exists and it is unique and linear.

\begin{figure}
\centering
\includegraphics [height=1.5 in, width=2 in]{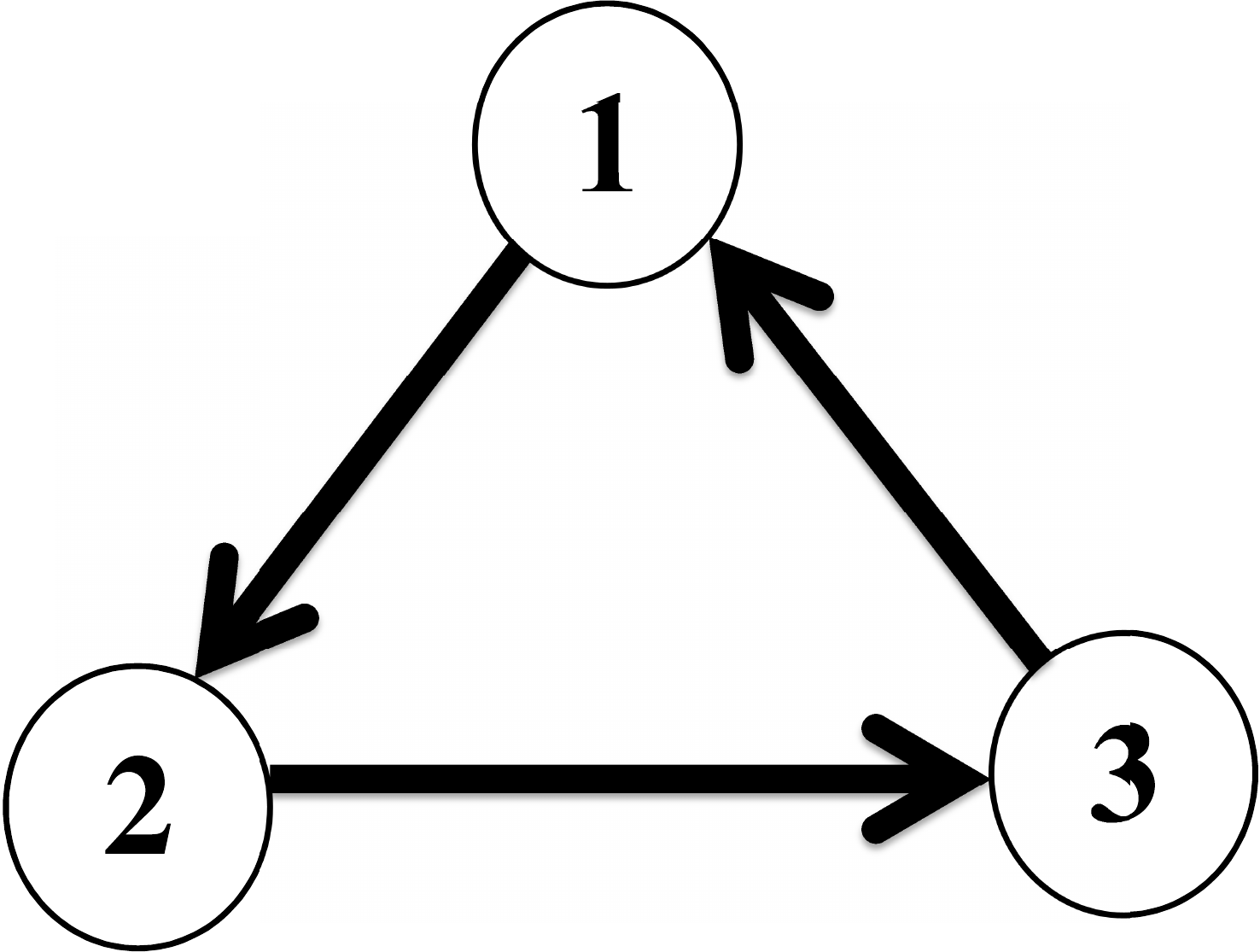}
\caption{The graph illustrates the interconnection structure of three players. The state of Player~$1$ at time $k+1$ depends directly on the state of Player~$3$ at time $k$ since $A_{13}\neq  0$, hence there is an arc from node $3$ to node $1$ in the interconnection graph. On the other hand, since $A_{12}=0$, Player~$1$ is not affected directly by the state of Player $2$, and there is no arc from node $2$ to node $1$ in the interconnection graph.} \label{fig:graph1}
\end{figure}

While the approach proposed in this paper applies for linear systems over strongly connected graphs, we will concentrate
on a simple delayed information control problem referred to as the three-player problem shown in Figure~\ref{fig:graph1}. For this problem, the system matrices have the structure
\begin{align*}
A=\begin{bmatrix} A_{11} &  0 & A_{13}\\  A_{21} & A_{22} & 0\\ 0 & A_{32} & A_{33} \end{bmatrix},\; B=\begin{bmatrix} B_{1} &  0 &  0\\ 0 & B_{2} &  0\\ 0 &  0 &  B_{3} \end {bmatrix},\; C=\begin{bmatrix} C_{1} &  0 &  0\\  0 & C_{2} &  0\\0 &  0 & C_{3} \end{bmatrix},
\end{align*}
and the information available to each player at time $k$ is
\begin{align*}
\mathcal{I}_{1}(k)&=\{y_1(k),y_1(k-1),y_3(k-1),y(0:k-2)\},\nonumber\\
\mathcal{I}_{2}(k)&=\{y_2(k),y_1(k-1),y_2(k-1),y(0:k-2)\},\\
\mathcal{I}_{3}(k)&=\{y_3(k),y_2(k-1),y_3(k-1),y(0:k-2)\}\nonumber.
\end{align*}
Since the information structure is partially nested, the optimal controller of each player is a linear function of the elements of its information set. Hence,
\begin{align*}
u_1(k)&=f_{11}\bigl(y_1(k)\bigr)+f_{12}\bigl(y_1(k-1),y_3(k-1)\bigr)+f_{13}\bigl(y(0:k-2)\bigr),\nonumber\\
u_2(k)&=f_{21}\bigl(y_2(k)\bigr)+f_{22}\bigl(y_1(k-1),y_2(k-1)\bigr)+f_{23}\bigl(y(0:k-2)\bigr),\nonumber\\
u_3(k)&=f_{31}\bigl(y_3(k)\bigr)+f_{32}\bigl(y_2(k-1),y_3(k-1)\bigr)+f_{33}\bigl(y(0:k-2)\bigr),
\end{align*}
where $f_{ij}$ is a linear function for all $i$, $j$. Therefore, $u(k)$ can be expressed as
\begin{eqnarray}
u(k)=F(k) y(k)+G(k){y(k-1)}+f\bigl({y}(0:k-2)\bigr),
\label{ucontrol}
\end{eqnarray}
where
\begin{align*}
f={\begin{bmatrix}  f_{13} \\ f_{23} \\ f_{33}  \end{bmatrix}},\
F(k)=\begin{bmatrix} F_{11}(k) &  0 &  0 \\  0 & F_{22}(k) &  0 \\  0 & 0 &  F_{33}(k)\\ \end{bmatrix},\;
G(k)=\begin{bmatrix} G_{11}(k) &  0 &  G_{13}(k) \\  G_{21}(k) &  G_{22}(k) &  0 \\  0 & G_{32}(k) & G_{33}(k)\\ \end{bmatrix}.
\end{align*}
Note that the sparsity structures of $F$ and $G$ comply with the information constraints at time $k$ and $k-1$, respectively. The control problem is now to find matrices $F$ and $G$, as well as a linear function $f$, that minimize $J$.

\section{Estimation Structure}
\label{sec:Estimation}

This section presents an optimal estimation algorithm for the three-player problem. First, we provide a short summary of standard Kalman filtering in Subsection~\ref{subsec:kalman}. Next, Subsection~\ref{subsec:kalman1} sketches a derivation of the estimation algorithm. Finally, some properties of the algorithm are given in Subsection~\ref{subsec:kalman2}.

\subsection{Preliminaries on Standard Kalman Filtering}
\label{subsec:kalman}

Consider a linear system on the form (\ref{augmentedsystem}), whose initial state $x(0)$ is Gaussian with zero mean and covariance matrix $P_0$. Let us define the following variables
\begin{eqnarray*}
\widehat{x}(k|k-1)& \defeq &\textbf{E}\{x(k)|y(0:k-1)\}\\
e(k)& \defeq &x(k)-\widehat{x}(k|k-1)\\
P(k)& \defeq &\textbf{E}\{e(k)e^T(k)\}.
\end{eqnarray*}
Here, $\widehat{x}(k|k-1)$ is the one-step prediction of the state, $e(k)$ is the prediction error, and $P(k)$ is the covariance matrix of the prediction error at time $k$. Assume that $u(k)$ is a deterministic function of $y(0:k)$. The Kalman filter equations can be written as follows (\cite{Astrom:70})
\begin{align}
\begin{split}
\widehat{x}(k+1|k)&= A\widehat{x}(k|k-1)+B u(k)+K(k)\bigl(y(k)-C\widehat{x}(k|k-1)\bigr)\label{201}\\
P(k+1) &= AP(k)A^T+W-AP(k)C^T\bigl(CP(k)C^T+V\bigr)^{-1}CP(k)A^T,
\end{split}
\end{align}
with $\widehat{x}(0|-1)=0$ and $P(0)=P_0$. Here, $K(k)$ is the optimal Kalman gain given by
\begin{align*}
K(k)&=AP(k)C^T\left(CP(k)C^T+V\right)^{-1}.
\end{align*}
The innovations are defined by
\begin{equation}
\widetilde{y}(k)=y(k)-C\widehat{x}(k|k-1).\label{905}
\end{equation}
The following proposition will be useful when deriving the optimal estimation algorithm for the three-player problem.
\begin{proposition}(\cite{Astrom:70})
The following facts hold:
\begin{enumerate}
\item[(a)] $\bold{{E}}\{x(k)\widetilde{y}(k)^T\}=P(k)C^T$.
\item[(b)] $\widetilde{y}(k)$ is an uncorrelated Gaussian process with covariance matrix
$
\widetilde{Y}(k)=CP(k)C^T+V
$. Moreover, under Assumption $1$, $\widetilde{Y}(k)$ is positive definite.
\item[(c)] $\widetilde{y}(k)$ is independent of past measurements
\begin{align*}
\bold{E}\{\widetilde{y}(k)y^T(j)\}=0\;\;\;\mbox{for}\;j<k.
\end{align*}
\end{enumerate}
\label{prop4}
\end{proposition}


\subsection{Kalman Filtering for Three-player Problem}
\label{subsec:kalman1}

Let $\mathcal{I}^{[1]}_{i}(k)$ be the set of all measurements up to time step $k-1$ that are available to Player $i$ at time $k$. For example,
\begin{align*}
\mathcal{I}^{[1]}_{1}(k)=\left\{y_1(k-1),y_3(k-1),y(0:k-2)\right\}.
\end{align*}
It is easy to verify that $\mathcal{I}^{[1]}_{i}(k)\subset y(0:k-1)$, \emph{i.e.} it does not have access to all measurements taken at time $k-1$. Hence, players cannot execute the one-step prediction of the standard Kalman filter $\widehat{x}_i(k|k-1)$ at time $k$. Define
\begin{align*}
\widehat{x}^{[1]}_i(k)\defeq \textbf{E}\left\{x_i(k)|\mathcal{I}^{[1]}_{i}(k)\right\},\;\;i=1,2,3.
\end{align*}
We will now derive explicit expressions for these quantities.

\begin{figure}
\centering
\includegraphics [height=2.5 in, width=3 in]{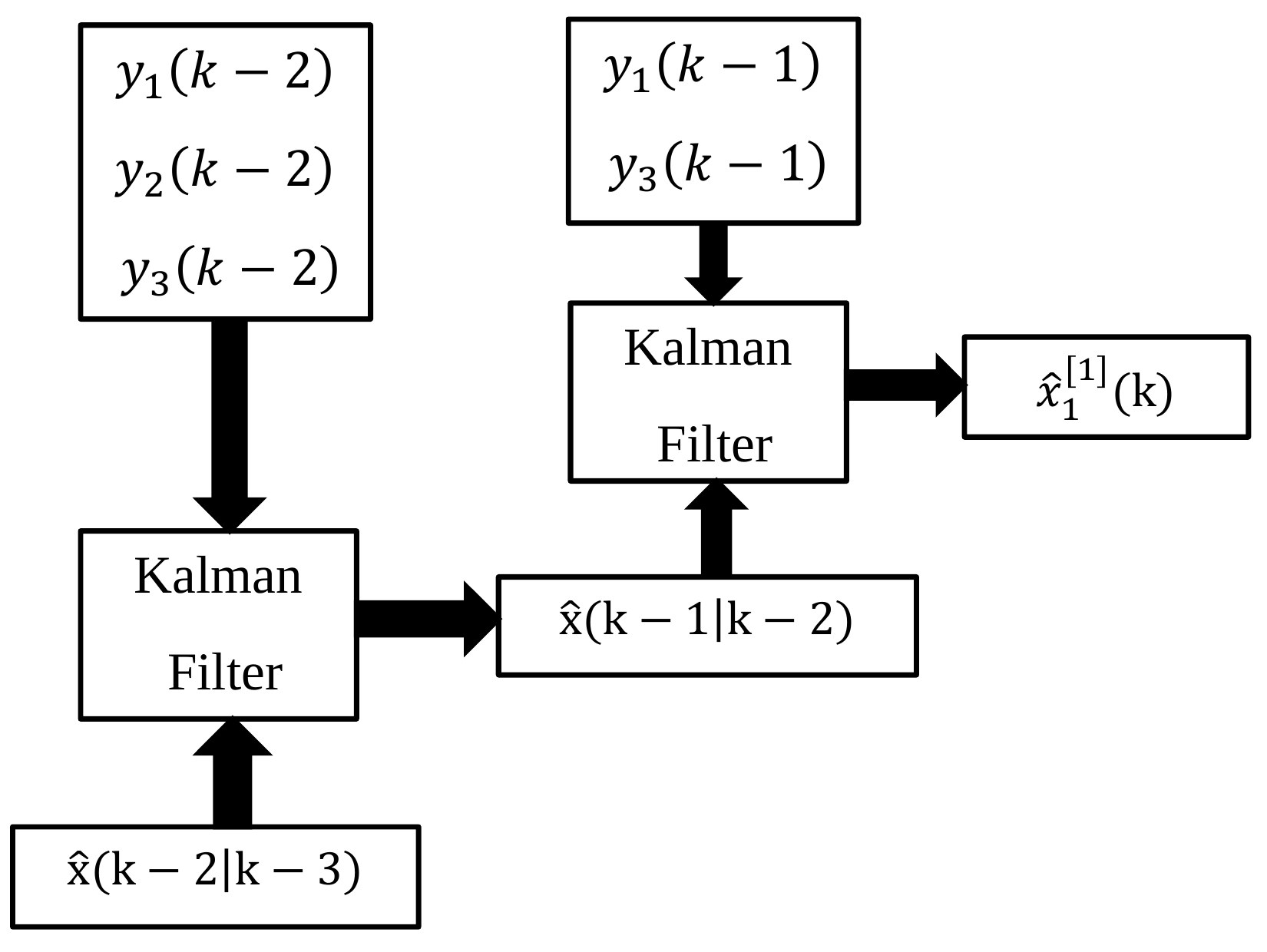}
\caption{Optimal estimation scheme of Player $\mbox{1}$ at time $k$.} \label{fig:graph2}
\end{figure}

Note that $y(0:k-2)$ is the piece of information available to all players. Thus, $\widehat{x}(k-1|k-2)$ can be computed by each player
at time $k$. To see how the optimal estimation algorithm for the three-player problem is derived, consider Player $1$. Let $[A]_i$ denote the $i$th block row of $A$. Then,
\begin{align*}
\widehat{x}^{[1]}_1(k)&= \textbf{E}\left\{x_1(k)|\mathcal{I}^{[1]}_{1}(k)\right\}\\
&=[A]_1\textbf{E}\left\{x(k-1)|\mathcal{I}^{[1]}_{1}(k)\right\}+B_1\textbf{E}\left\{u_1(k-1)|\mathcal{I}^{[1]}_{1}(k)\right\}\\
&=[A]_1\textbf{E}\left\{x(k-1)|y_1(k-1),y_3(k-1),y(0:k-2)\right\}+B_1u_1(k-1),
\end{align*}
where we used the independence of $w_1(k-1)$ and $\mathcal{I}^{[1]}_{1}(k)$, and the fact that $u_1(k-1)$ is a deterministic function of the information set $\mathcal{I}^{[1]}_{1}(k)$. To evaluate the expected value of $x(k-1)$ given $\mathcal{I}^{[1]}_{1}(k)$, we will first change the variables so that we get independent variables. According to Proposition $1(\mbox{c})$, the innovations $\widetilde{y}_1(k-1)$ and $\widetilde{y}_3(k-1)$ are independent of $y(0:k-2)$. Thus,
\begin{align}
\widehat{x}^{[1]}_1(k)=&[A]_1\textbf{E}\{x(k-1)|y(0:k-2)\}+[A]_1\textbf{E}\{x(k-1)|\widetilde{y}_1(k-1),\widetilde{y}_3(k-1)\}+B_1u_1(k-1)\nonumber\\
=&[A]_1\widehat{x}(k-1|k-2)+B_1u_1(k-1)+[A]_1\textbf{E}\{x(k-1)|\widetilde{y}_1(k-1),\widetilde{y}_3(k-1)\},\label{1000}
\end{align}
where we used Proposition $\mbox{4}(\mbox{a})$ to get the first equality. We will now calculate the last term of Equation \eqref{1000}. Let $S_t=\{1,2,3\}$ and $S_1=\{1,3\}$. Then
\small
\begin{align}
\textbf{E}\{x(k-1)&|\widetilde{y}_1(k-1),\widetilde{y}_3(k-1)\}\nonumber\\
&=\textbf{{Cov}}\left\{x(k-1),\begin{bmatrix}\widetilde{y}_1(k-1)\\\widetilde{y}_3(k-1)
\end{bmatrix}\right\}
\textbf{{Cov}}^{-1}\left\{\begin{bmatrix}\widetilde{y}_1(k-1)\\\widetilde{y}_3(k-1)\end{bmatrix},
\begin{bmatrix}\widetilde{y}_1(k-1)\\\widetilde{y}_3(k-1)\end{bmatrix}\right\}
\begin{bmatrix}\widetilde{y}_1(k-1)\\\widetilde{y}_3(k-1)\end{bmatrix}\nonumber\\
&\hspace{-2cm}=\begin{bmatrix}[P(k-1)]_{11}C_1^T & [P(k-1)]_{13}C_3^T \\ [P(k-1)]_{21}C_1^T & [ P(k-1)]_{23}C_3^T\\ [ P(k-1)]_{31} C_1^T & [ P(k-1)]_{33}  C_3^T\end{bmatrix}\begin{bmatrix}  C_1[ P(k-1)]_{11} C_1^T+[V]_{11}&  C_1[ P(k-1)]_{13} C_3^T+[V]_{13}\nonumber\\
C_3[P(k-1)]_{31} C_1^T+[V]_{31} &  C_3[ P(k-1)]_{33} C_3^T+[V]_{33} \end{bmatrix}^{-1}\begin{bmatrix}\widetilde{y}_1(k-1)\\\widetilde{y}_3(k-1)\end{bmatrix}\\
&=[P(k-1)]_{S_tS_1}[C]_{S_1S_1}^T\bigl([C]_{S_1S_1} [P(k-1)]_{S_1S_1}[C]_{S_1S_1}^T+ [V]_{S_1S_1}\bigr)^{-1}\begin{bmatrix}\widetilde{y}_1(k-1)\\\widetilde{y}_3(k-1)\end{bmatrix},\label{1001}
\end{align}
\normalsize
where we used Proposition $\mbox{4}(\mbox{b})$ to get the first equality and Proposition $\mbox{1}(\mbox{a})$-$\mbox(\mbox{b})$ to obtain the second equality. Substituting Equation \eqref{1001} into Equation \eqref{1000} shows that $\widehat{x}^{[1]}_1(k)$ is computed as
\begin{align}
\widehat{x}^{[1]}_1(k)=&[A]_1\widehat{x}(k-1|k-2)+B_1u_1(k-1)+\begin{bmatrix}{K}^{[1]}_{11}(k-1) & {K}^{[1]}_{13}(k-1)\end{bmatrix}
\begin{bmatrix}\widetilde{y}_1(k-1)\\\widetilde{y}_3(k-1)\end{bmatrix},\label{1002}
\end{align}
where
\small
\begin{align*}
\begin{bmatrix}{K}^{[1]}_{11}(k-1) & {K}^{[1]}_{13}(k-1)\end{bmatrix}=[A]_1[P(k-1)]_{S_tS_1}[C]_{S_1S_1}^T\bigl ([C]_{S_1S_1} [P(k-1)]_{S_1S_1}[C]_{S_1S_1}^T+ [V]_{S_1S_1}\bigr)^{-1}.
\end{align*}
\normalsize
Similar results can be obtained for Player $2$ and Player $3$. Let $S_2=\{1,2\}$ and $S_3=\{2,3\}$. Then
\begin{align}
\widehat{x}^{[1]}_2(k)=&[A]_2\widehat{x}(k-1|k-2)+B_2u_2(k-1)+\begin{bmatrix}{K}^{[1]}_{21}(k-1) & {K}^{[1]}_{22}(k-1)\end{bmatrix}
\begin{bmatrix}\widetilde{y}_1(k-1)\\\widetilde{y}_2(k-1)\end{bmatrix},\label{1003}\\
\widehat{x}^{[1]}_3(k)=&[A]_3\widehat{x}(k-1|k-2)+B_3u_3(k-1)+\begin{bmatrix}{K}^{[1]}_{32}(k-1) & {K}^{[1]}_{33}(k-1)\end{bmatrix}
\begin{bmatrix}\widetilde{y}_2(k-1)\\\widetilde{y}_3(k-1)\end{bmatrix},\label{1004}
\end{align}
where
\small
\begin{align*}
\begin{bmatrix}{K}^{[1]}_{21}(k-1) & {K}^{[1]}_{22}(k-1)\end{bmatrix}&=[A]_2[P(k-1)]_{S_tS_2}[C]_{S_2S_2}^T\left ([C]_{S_2S_2} [P(k-1)]_{S_2S_2}[C]_{S_2S_2}^T+ [V]_{S_2S_2}\right)^{-1},\\
\begin{bmatrix}{K}^{[1]}_{32}(k-1) & {K}^{[1]}_{33}(k-1)\end{bmatrix}&=[A]_3[P(k-1)]_{S_tS_3}[C]_{S_3S_3}^T\left ([C]_{S_3S_3} [P(k-1)]_{S_3S_3}[C]_{S_3S_3}^T+ [V]_{S_3S_3}\right)^{-1}.
\end{align*}
\normalsize
Define the matrix ${K}^{[1]}$ by
\begin{align*}
{K}^{[1]}(k)=\begin{bmatrix} {K}^{[1]}_{11}(k) &  0 & {K}^{[1]}_{13}(k)\vspace{1mm}\\
{K}^{[1]}_{21}(k) & {K}^{[1]}_{22}(k) &  0\vspace{1mm}\\  0 & {K}^{[1]}_{32}(k) & {K}^{[1]}_{33}(k)\end{bmatrix}.
\end{align*}
Then equations \eqref{1002}-\eqref{1004} can be combined and written in the compact form as
\begin{eqnarray}
\widehat{x}^{[1]}(k)&=A\widehat{x}(k-1|k-2)+B u(k-1)+{K}^{[1]}(k-1)\bigl({y}(k-1)-C\widehat{x}(k-1|k-2)\bigr).
\label{200}
\end{eqnarray}
The Kalman filter iterations for the three-player problem at time $k$ is summarized as follows
\boxedeqn{
\widehat{x}(k-1|k-2)&=A\widehat{x}(k-2|k-3)+B u(k-2)+{K}(k-2)\bigl({y}(k-2)-C\widehat{x}(k-2|k-3)\bigr)\nonumber\\
\widehat{x}^{[1]}(k)&=A\widehat{x}(k-1|k-2)+B u(k-1)+{K}^{[1]}(k-1)\bigl({y}(k-1)-C\widehat{x}(k-1|k-2)\bigr).\label{itkal}}
Note that ${K}^{[1]}$ is not the usual Kalman filter gain and that it has a the same sparsity pattern as
$G$. Figure \ref{fig:graph2} shows the overall estimation scheme of Player $\mbox{1}$ at time $k$.

\begin{remark} Both ${K}^{[1]}$ and $K$ can be calculated off-line without knowing the control
input history $u(0:N-1)$.
\end{remark}


\subsection{Estimator properties}
\label{subsec:kalman2}
Here we compute some quantities that will help us in the following section. Define
\begin{align*}
{e}^{[1]}(k)&\defeq x(k)-\widehat{x}^{[1]}(k)\\
\widetilde{y}^{\;[1]}(k)&\defeq y(k)-C\widehat{x}^{[1]}(k).
\end{align*}
We denote the covariance matrices of ${e}^{[1]}(k)$ and $\widetilde{y}^{\;[1]}(k)$ by $P^{[1]}(k)$ and $\widetilde{Y}^{[1]}(k)$, respectively.
\begin{lemma}
Let $\bigtriangleup K(k)=K(k)-K^{[1]}(k)$. Then the following facts hold:
\begin{enumerate}
\item[(a)] $P^{[1]}(k)=P(k)+\bigtriangleup K(k-1)\widetilde{Y}(k-1)\bigtriangleup K^T(k-1).$
\item[(b)] $\widetilde{Y}^{[1]}(k)=CP^{[1]}(k)C^T+V$. Also, under Assumption~$\mbox{1}$, $\widetilde{Y}^{[1]}(k)$ is positive definite.
\item[(c)] $\widetilde{P}(k)\defeq\textbf{E}\left\{{e}^{[1]}(k)\widetilde{y}^{\;T}(k-1)\right\}=
               \bigtriangleup K(k-1)\widetilde{Y}(k-1)$.
\end{enumerate}
\label{lemma1}
\end{lemma}
\begin{IEEEproof} See Appendix.
\end{IEEEproof}


\section{Optimal Controller Derivation}
\label{sec:Decomposition}

This section shows that finding optimal controller for the three-player problem is equivalent to solving two separate optimization
problems. Before proceeding, we state the following proposition.
\begin{proposition} (\cite {Astrom:70})
Define the matrices
\small
\begin{align}
S(k)&=A^TS(k+1)A+Q_{xx}-\bigl(A^TS(k+1)B+Q_{xu}\bigr)\bigl(B^T S(k+1) B+Q_{uu}\bigr)^{-1}\bigl(B^T S(k+1)A+Q^T_{xu}\bigr)\nonumber\\
H(k)&=B^T S(k+1) B+Q_{uu}\label{lqg}\\
L(k)&={H^{-1}(k)}\bigl(B^T S(k+1) A+ Q_{xu}^T\bigr)\nonumber,
\end{align}
\normalsize
for $k=0,\cdots,N-1$ and where $S(N)=Q_0$. Then the cost function (\ref{cost}) can be written as
\begin{align*}
J=&\underbrace{\sum_{k=0}^{N-1}\bold{E}\left\{\bigl(u(k)-L(k)x(k)\bigr)^TH(k)\bigl(u(k)-L(k)x(k)\bigr)\right\}}_{J_u}\\
&+\underbrace{\bold{Tr}\{S(0)P_0\}+\sum_{k=0}^{N-1}\bold{Tr}\{S(k+1)W\}}_{J_w}.
\end{align*}
Moreover, $J_{w}$ is independent of the control.
\label{dec}
\end{proposition}
From Proposition \ref{dec}, it can be seen that minimizing $J$ is equivalent to minimizing $J_u$. Also, under Assumption~\ref{costmatrix}, $H(k)$ is positive definite for all $k$.

The first step towards finding the structure of the optimal controller is to decompose the state vector into independent terms using the following lemma:
\begin{lemma}\label{lemma2}
The state vector can be decomposed
as
\begin{align*}
x(k)&=\widetilde{x}(k)+\widehat{x}(k),
\end{align*}
where $\widehat{x}(k)$ and $\widetilde{x}(k)$ are independent and given by
\begin{align*}
\widehat{x}(k)&=\bold{E}\{x(k)|y(0:k-2)\}\\
\widetilde{x}(k)&={e}^{[1]}(k)+\bigl(BF(k-1)+{K}^{[1]}(k-1)\bigr)\widetilde{y}(k-1).
\end{align*}
\end{lemma}
\begin{IEEEproof} See appendix.
\end{IEEEproof}

Note that the term $\widehat{x}(k)$ is the conditional estimate of the state $x(k)$ given the  information shared by all players, and $\widetilde{x}(k)$ is the estimation error. Now that the state vector has been decomposed into independent terms, the control input $u(k)$ can be decomposed in an analogue manner.
\begin{lemma}\label{lemma3}
The control input $u(k)$ can be decomposed into two independent terms
\begin{align*}
u(k)=\widetilde{u}(k)+\widehat{u}(k),
\end{align*}
where
\begin{align*}
\widehat{u}(k)&=\bold{E}\{u(k)|y(0:k-2)\}\\
\widetilde{u}(k)&=F(k)\widetilde{y}^{[1]}(k)+F^{[1]}(k)\widetilde{y}(k-1),
\end{align*}
and $F^{[1]}$ is given by
\begin{equation}
F^{[1]}(k)=G(k)+F(k)C\bigl({K}^{[1]}(k-1)+BF(k-1)\bigr).\label{G}
\end{equation}
\end{lemma}
\begin{IEEEproof} See appendix.
\end{IEEEproof}

\begin{remark}
Since $B$, $C$ and $F$ are diagonal matrices, $G(k)$ and $F(k)C{K}^{[1]}(k-1)$ have the same sparsity pattern. Similarly, $F^{[1]}(k)$ and $G(k)$ have the same sparsity pattern.
\end{remark}

From lemmas \ref{lemma2} and \ref{lemma3}, both $\widehat{x}(k)$ and $\widehat{u}(k)$ are functions of $y(0:k-2)$ which is independent of $\widetilde{x}(k)$ and $\widetilde{u}(k)$.
As a result the cost function $J_u$ can be decomposed as
\begin{align*}
J_u&=\underbrace{\sum_{k=0}^{N-1}\textbf{E}\left\{\bigl(\widetilde{u}(k)-L(k)\widetilde{x}(k)\bigr)^T H(k)\bigl(\widetilde{u}(k)-L(k)\widetilde{x}(k)\bigr)\right\}}_{\widetilde{J}}\nonumber\\
&+\underbrace{\sum_{k=0}^{N-1}\textbf{E}\left\{\bigl(\widehat{u}(k)-L(k)\widehat{x}(k)\bigr)^T H(k)\bigl(\widehat{u}(k)-L(k)\widehat{x}(k)\bigr)\right\}}_{\widehat{J}},\nonumber
\end{align*}
and the optimal control problem reduces to solving
\begin{align*}
\hspace{-5.0cm}\;\;\;\;\;\;\;\mbox{\textbf{Problem 1.}}\;\;\mbox{minimize}&\;\;\;\widehat{J}(\widehat{x},\widehat{u})\nonumber\\
\mbox{subject to}&\;\;\;\widehat{u}(k)\;\mbox{is a function of}\;y(0:k-2).
\end{align*}
\begin{align*}
\hspace{0.8cm}\;\;\mbox{\textbf{Problem 2.}}\;\;\mbox{minimize}&\;\;\;\widetilde{J}(\widetilde{x},\widetilde{u})\nonumber\\
\mbox{subject to}&\;\;\; \widetilde{u}(k)=F(k) \widetilde{y}^{[1]}(k)+F^{[1]}(k) \widetilde{y}(k-1),\\
&\;\;\; F(k)\; \mbox{and}\; F^{[1]}(k)\; \mbox{have specified sparsity structures.}
\end{align*}

The following lemma shows that the optimal solution $\widehat{u}(k)$ for Problem $\mbox{1}$  is exactly the optimal controller for centralized information structure with two-step delay, where the information set of each player is $y(0:k-2)$.
\begin{lemma}
Suppose assumptions $1$ and $2$ hold. An optimal solution for Problem $\mbox{1}$ is given by
\begin{eqnarray}
\widehat{u}(k)&=&L(k)\widehat{x}(k)\nonumber\\
&=&L(k)\bold{E}\{x(k)|y(0:k-2)\}.
\label{500}
\end{eqnarray}
Moreover, the optimal value of the cost function $\widehat{J}$ is zero.
\end{lemma}
\begin{IEEEproof} See appendix.
\end{IEEEproof}

We now focus on Problem $\mbox{2}$, namely the computation of $\left\{F(k)\right\}_{k=0,\ldots,N-1}$ and $\left\{F^{[1]}(k)\right\}_{k=1,\ldots,N-1}$. Recalling the expansions of $\widetilde{x}(k)$ and $\widetilde{u}(k)$ in terms of $\widetilde{y}^{[1]}(k)$, ${e}^{[1]}(k)$, and $\widetilde{y}(k-1)$, $\widetilde{J}$ can be expanded as follows
\small
\begin{align}
\widetilde{J}=&\sum_{k=0}^{N-1}\textbf{E}\left\{\bigl(\widetilde{u}(k)-L(k)\widetilde{x}(k)\bigr)^T H(k)\bigl(\widetilde{u}(k)-L(k)\widetilde{x}(k)\bigr)\right\}\nonumber\\
=&\sum_{k=0}^{N-1}\textbf{Tr}\left\{H(k)F(k)VF^T(k)\right\}+\textbf{Tr}\left\{H(k)\bigl(F(k)C-L(k)\bigr)P^{[1]}(k)\bigl(F(k)C-L(k)\bigr)^T\right\}\nonumber\\
&+\textbf{Tr}\left\{H(k)\bigl(F^{[1]}(k)-L(k)(BF(k-1)+{K}^{[1]}(k-1)\bigr)\widetilde{Y}(k-1)
\bigl(F^{[1]}(k)-L(k)(BF(k-1)+{K}^{[1]}(k-1)\bigr)^T\right\}\nonumber\\
&+2\textbf{Tr}\left\{H(k)\bigl(F(k)C-L(k)\bigr)\widetilde{P}(k)
\bigl(F^{[1]}(k)-L(k)(BF(k-1)+{K}^{[1]}(k-1)\bigr)^T\right\}\label{370},
\end{align}
\normalsize
where we used Proposition $\mbox{4(c)}$. A point worth noticing is that according to Proposition $\mbox{1}$ and Lemma $\mbox{1}$, $P^{[1]}$, $\widetilde{P}$, and  $\widetilde{Y}$ are independent of $F(k)$ and $F^{[1]}(k)$. To minimize $\widetilde{J}$ with respect to $F(k)$ and $F^{[1]}(k)$, we face two difficulties: the first is that $F(k)$ and $F^{[1]}(k)$ must satisfy given sparsity constraints; the second difficulty is the existence of coupling terms between $F(k-1)$ and $F(k)$. To overcome these difficulties, we will use the vec operator and the following lemma:
\begin{lemma}
Assume that $A \in\mathbb{R}^{n\times m}$ is split into sub-blocks as follows:
\begin{align*}
A=\begin{bmatrix} A_{11} & \cdots & A_{1q}\\ \vdots &  & \vdots\\ A_{p1} & \cdots & A_{pq}\end{bmatrix},
\end{align*}
where $A \in\mathbb{R}^{n_i\times m_j}$ for $i=1,\ldots,p$ and $j=1,\ldots,q$. Let $S$ be the set of non-zero
sub-blocks of $A$,
$$
S=\{A_{ij}\mid A_{ij}\neq 0\},\; \mid S\mid= s.
$$
Then there always exists a full column rank matrix $E$ of an appropriate dimension such that
$$
\textup{vec}(A)=E \begin{bmatrix} \textup{vec}(A_{i_1j_1}) \\  \vdots\\ \textup{vec}(A_{i_sj_s})\end{bmatrix},
$$
where $A_{i_kj_k} \in S$ for all $k=1,\ldots,s$.
\end{lemma}
\begin{IEEEproof} See appendix.
\end{IEEEproof}
The way to construct matrix $E$ is described in Appendix. Lemma $\mbox{5}$ ensures the existence of $E_1$ and $E_2$ such that
\begin{align*}
\textup{vec}\bigl(F(k)\bigr)&=E_1\xi_1(k),\\
\textup{vec}\left(F^{[1]}(k)\right)&=E_2\xi_2(k),
\end{align*}
where $\xi_1$ and $\xi_2$ are vectors formed by stacking all nonzero sub-blocks of $F$ and $F^{[1]}$, respectively. That is,
\small
\begin{align*}
\textrm{vec}\bigl(F(k)\bigr)&=E_1 \underbrace{\begin{bmatrix} \textrm{vec}^T\left(F_{11}\right) & \textrm{vec}^T\left(F_{22}\right) & \textrm{vec}^T\left(F_{33}\right) \end{bmatrix}^T}_{\xi_1(k)},\\
\textrm{vec}\left(F^{[1]}(k)\right)&=E_2 \underbrace{\begin{bmatrix} \textrm{vec}^T\left(F^{[1]}_{11}\right) & \textrm{vec}^T\left(F^{[1]}_{21}\right) & \textrm{vec}^T\left(F^{[1]}_{22}\right) & \textrm{vec}^T\left(F^{[1]}_{32}\right) & \textrm{vec}^T\left(F^{[1]}_{13}\right) & \textrm{vec}^T\left(F^{[1]}_{33}\right)
\end{bmatrix}^T}_{\xi_2(k)}.
\end{align*}
\normalsize
We now show how vectorization allows to convert Problem $\mbox{2}$ into an unconstrained convex optimization problem.
\begin{lemma}
Let $E=\bold{diag}(E_1,E_2)$, $\zeta(k)= \begin{bmatrix} \xi_1(k-1)\\ \xi_2(k)\end{bmatrix}$ for $k=1,\ldots,N-1$, and
$\zeta(N)=\xi_1(N-1)$. Define
\small
\begin{align*}
Z_{1}(k)&=E^T\begin{bmatrix} I & 0 \\ -I \otimes L(k)B & I \end{bmatrix}^T\begin{bmatrix} \widetilde{Y}^{[1]}(k-1)\otimes H(k-1)
 & 0 \\  0 & \widetilde{Y}(k-1)\otimes H(k) \end{bmatrix}
\begin{bmatrix} I & 0 \\ -I \otimes L(k)B & I \end{bmatrix}E,\\
Z_{2}(k)&=E\begin{bmatrix} -I \otimes L(k)B & I \end{bmatrix} ^T \left(\widetilde{P}^T(k)C^T\otimes H(k)\right)\begin{bmatrix} I & 0 \end{bmatrix}E,\\
b(k)&=E^T\begin{bmatrix} I & 0 \end{bmatrix}^T \left(CP^{[1]}(k-1)\otimes H(k-1)\right)\textup{vec}\bigl(L(k-1)\bigr)\\
&+E^T\begin{bmatrix} -I \otimes L(k)B & I \end{bmatrix}^T \left(\widetilde{Y}(k-1)\otimes H(k) \right)\textup{vec}\bigl(L(k)K^{[1]}(k-1)\bigr),
\end{align*}
\normalsize
with
\small
\begin{align*}
Z_1(N)=&E_1^T\bigl(\widetilde{Y}^{[1]}(N-1)\otimes H(N-1)\bigr)E,\\
b(N)=&E_1^T\left(CP^{[1]}(N-1)\otimes H(N-1)\right)\textup{vec}\bigl(L(N-1)\bigr).
\end{align*}
\normalsize
Then Problem~$\mbox{2}$ is equivalent to
\small
\begin{align}
\min_{\zeta(1),\ldots,\zeta(N)}=&\sum_{k=1}^{N-1}{1 \over 2}\zeta^T(k){Z}_1(k)\zeta(k)+\zeta^T(k){Z}_2(k)\zeta(k+1)
-\zeta^T(k){b}(k)\nonumber\\
&+{1 \over 2}\zeta^T(N){Z}_1(N)\zeta(N)-\zeta^T(N){b}(N)\label{59}
\end{align}
\normalsize
 Moreover, ${Z}_1(k)$ is positive definite for all $k$.
\end{lemma}
\begin{IEEEproof} See Appendix.
\end{IEEEproof}
Consider the two time-step case of~(\ref{59})
\begin{align}
\min_{\zeta(1),\zeta(2)}\underbrace{{1 \over 2}\zeta^T(1){Z}_1(1)\zeta(1)-\zeta^T(1){b}(1)}_{g_1(\zeta(1))}+\underbrace{\zeta^T(1){Z}_2(1)\zeta(2)+{1 \over 2}\zeta^T(2){Z}_1(2)\zeta(2)-\zeta^T(2){b}(2)}_{g_2(\zeta(1),\zeta(2))}.\label{example}
\end{align}
The optimal $\zeta(2)$ is the one which minimizes $g_2$, \emph{i.e.}
\begin{align*}
\zeta^{\star}(2)&=\argmin_{\zeta(2)}g_2\bigl(\zeta(1),\zeta(2)\bigr)\\
&=-Z_1^{-1}(2)\bigl({Z}^T_2(1) \zeta(1)-b(2)\bigr).
\end{align*}
If we substitute the optimal $\zeta^{\star}(2)$ into \eqref{example}, then we can minimize $g_1\bigl(\zeta(1)\bigr)+g_2\bigl(\zeta(1),\zeta^{\star}(2)\bigr)$ with respect to $\zeta(1)$. Therefore,
\begin{align*}
\zeta^{\star}(1)&=\argmin_{\zeta(1)}g_1\bigl(\zeta(1)\bigr)+g_2\bigl(\zeta(1),\zeta^{\star}(2)\bigr)\\
&=R_1^{-1}(1)c(1),
\end{align*}
where
\begin{eqnarray*}
R(1)&=&{Z}_1(1)-{Z}_2(1)Z_1^{-1}(2){Z}^T_2(1),\\
c(1)&=&{b}(1)-{Z}_2(1)Z_1^{-1}(2)b(2).
\end{eqnarray*}
The extension to more time steps is straightforward. The result is stated in the following lemma.
\begin{lemma}
Suppose assumptions $1$ and $2$ hold. Define
\begin{eqnarray*}
R(k)&=&{Z}_1(k)-{Z}_2(k)R^{-1}(k+1){Z}^T_2(k)\\
c(k)&=&{b}(k)-{Z}_2(k)R^{-1}(k+1)c(k+1),
\end{eqnarray*}
with the end condition $R(N)={Z}_1(N)$ and $c(N)=b(N)$. Then optimization problem~(\ref{59}) has the unique solution
\begin{align}
\zeta(k+1)=-R^{-1}(k+1)\bigl({Z}^T_2(k) \zeta(k)-c(k+1)\bigr),\label{gain}
\end{align}
with initial condition $\zeta(1)=R^{-1}(1)c(1)$. Moreover, ${R}(k)$ is positive definite for all $k$.
\end{lemma}

\section{Main Results}
\label{sec:Decomposition1}

We can now state our main result, Theorem $\mbox{1}$, which gives the optimal controller for the three-player problem.
\begin{theorem}
Suppose assumptions $1$ and $2$ hold. Let $\hat{x}(k)=\bold{E}\{x(k)|y(0:k-2)\}$. Then optimal controller for the three-player problem is given by
\begin{align}
u(k) &= F(k)\bigl(y(k)-C\widehat{x}^{[1]}(k)\bigr)+F^{[1]}(k)\bigl(y(k-1)-C\hat{x}(k-1\vert k-2) \bigr)+ L(k)\hat{x}(k),\label{ucontrol2}
\end{align}
where $\widehat{x}^{[1]}(k)$ and $\hat{x}(k-1\vert k-2)$ are the optimal state estimates obtained using the Kalman filter iterations \eqref{itkal}, $L$ is given by Equation (\ref{lqg}), and $F$ and $F^{[1]}$ are given by Equation (\ref{gain}). Moreover,
\begin{align*}
\hat{x}(k)=&\widehat{x}^{[1]}(k)-\bigl(BF(k-1)+{K}^{[1]}(k-1)\bigr)\bigl(y(k-1)-C\hat{x}(k-1\vert k-2) \bigr).
\end{align*}
\end{theorem}
Having derived the optimal controller, a number of remarks are in order.
\begin{remark}
A physical interpretation of the optimal control policy is given as follows: The third term of optimal controller, $L(k)\hat{x}(k)$, is exactly the optimal policy for centralized information structure with two-step delay, where the information set of each player is $y(0:k-2)$. The first and second terms are correction terms based on local measurements from time $k$ and $k-1$, respectively, which are available to each player.
\end{remark}
\begin{remark}
The recursive equation \eqref{gain} reveals a new feature present neither in LQG control with one-step delay sharing information pattern nor in the state-feedback case: the optimal control gain at time $k$, $\zeta(k)$, is an affine function of $\zeta(k-1)$. For example, in the state-feedback case where $y_i(k)=x_i(k)$ for $i=1,2,3$, we have
\begin{align*}
\widetilde{P}(k)=\mathbf{E}\{w(k-1)w^T(k-2)\}=0.
\end{align*}
According to Lemma $\mbox{6}$, ${Z}_2(k)=0$, and hence Equation~\eqref{gain} reduces to
$$\zeta(k)=Z_1^{-1}(k)b(k).$$
\end{remark}
\begin{remark}
Equating the right hand side of equations \eqref{ucontrol} and \eqref{ucontrol2} shows that the linear function $f$ is given by
\begin{align*}
f=\bigl(L(k)-F(k)C\bigr)\hat{x}(k)-G(k)C\hat{x}(k-1\vert k-2),
\end{align*}
where $G$ is given by Equation \eqref{G}. Note that both $\hat{x}(k)$ and $\hat{x}(k-1\vert k-2)$ are linear functions of $y(0:k-2)$.
\end{remark}
\begin{remark}
If $A \in\mathbb{R}^{n\times n}$, then the optimal controller for the three-player problem has at most $2n$ states.
\end{remark}

\section{Numerical Example}
\label{sec:Example}

We conclude our discussion of the three-player problem with an example. Consider a simple system specified by
\begin{align*}
A=\begin{bmatrix} 2 &  0 &  1\\  1 & 2 &  0\\ 0 &  1 &  2 \end{bmatrix},\;
B=\begin{bmatrix} 1 &  0 &  0\\  0 & 1 &  0\\ 0 &  0 &  1 \end{bmatrix},\;
C=\begin{bmatrix} 1 &  0 &  0\\  0 & 1 &  0\\ 0 &  0 &  1 \end{bmatrix}.
\end{align*}
$w$ and $v$ are Gaussian with zero mean and identity covariance matrix. The time horizon ${N}$ is chosen to be $1000$ and the
cost weight matrices are given by
\begin{align*}
Q_{xx}=\begin{bmatrix} 3 &  1 &  1\\  1 & 3 &  1\\ 1 &  1 &  3 \end{bmatrix},
Q_{xu}=\begin{bmatrix} 1 &  0 &  -1\\  -1 & 1 &  0\\ 0 &  -1 &  1 \end{bmatrix},
Q_{uu}=\begin{bmatrix} 2 &  0 &  0\\  0 & 2 &  0\\ 0 &  0 &  2 \end{bmatrix},
\end{align*}
and $Q_0=Q_{xx}$.

\begin{figure}
\centering
\includegraphics [height=0.8 in, width=0.8 in]{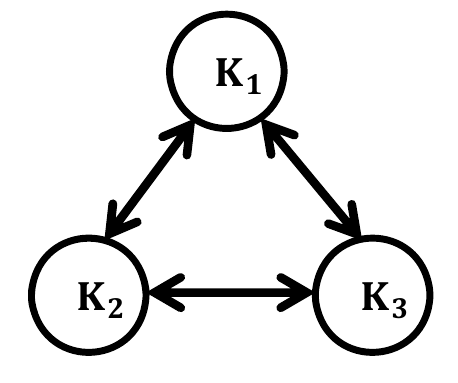}
\caption{The graph illustrates the communication structure of one-step delay information pattern. Each controller passes information to both neighbors after one-step delay.} \label{fig:graph3}
\end{figure}

We will compare the optimal controller for the three-player problem to controllers for the following information structures
\begin{enumerate}
\item Centralized with two-step delay: $u_i(k)=\mu_i\bigl(y(0:k-2)\bigr)$,
\item One-step delay sharing information pattern: $u_i(k)=\mu_i\bigl(y_i(k),y(0:k-1)\bigr)$,
\item Centralized without delay: $u_i(k)=\mu_i\bigl(y(0:k)\bigr)$.
\end{enumerate}
The one-step delay sharing information pattern studied in \cite {Sandell:74,Kurtaran:74,Toda:75,Yoshi:75} is specified by the graph in Figure~\ref{fig:graph3}.

By minimizing cost function \eqref{cost}, we obtain Table $\mbox{1}$. Centralized controller without delay has the lowest cost as expected. The three-player controller outperforms the centralized controller with two-step delay by a substantial margin, and only around $\mbox{1.74}\%$ higher than one-step delay sharing information pattern control. In other words, for three-player problem, there is a slight benefit of having two-way communication between controllers.
\begin{table}[height=4 in, width=4 in]
\centering
\caption{Simulation Results for Total Cost}
\begin{tabular}{|c | c |}
\hline
Control law & Cost mean \\
\hline
Centralized with delay& 14757 \\
\hline
Three-player & 339.9 \\
\hline
One-step delay information pattern& 334.1 \\
\hline
Centralized without delay& 188.8\\
\hline
\end{tabular}
\end{table}
Comparison of the costs shows the benefits of using all available information.


\section{Conclusion}
\label{sec:Conclusion}

In this paper, we presented an explicit solution for a distributed LQG problem in which three players communicate their information with delays. This was accomplished via decomposition of the state and input vectors into two independent terms and using this decomposition to
separate the optimal control problem to two subproblems. Computing the gains of the optimal controller requires solving one standard discrete-time Riccati equation and one recursive equation. Future work will continue to extend our approach to the infinite-horizon and more general networks.


\bibliography{interference}
\bibliographystyle{IEEEtr}


\section{Appendix}
\subsection{Preliminaries}
\begin{proposition}~(\cite{Horn:96}) If $A$, $B$, $C$, $D$, $X$ and $Y$ are suitably dimensioned matrices, then
\begin{enumerate}
\item[(a)] $\textup{vec}(AXB)=(B^T\otimes A)\textup{vec}(X)$,
\item[(b)] If $A$ and $B$ are positive definite, then so is $A\otimes B$,
\item[(c)] $\bold{Tr}\{AXBY^T\}=\textup{vec}^T(Y)(B^T\otimes  A)\textup{vec}(X)$,
\item[(d)] $(A\otimes B)^{-1}=A^{-1}\otimes B^{-1}$.
\item[(e)] Let $X\in\mathbb{R}^{m\times n}$, then there exists a unique permutation matrix $P_{m,n}\in\mathbb{R}^{mn\times mn}$ such that
$\textup{vec}(X^T)=P_{m,n}\textup{vec}(X)$. The matrix $P_{m,n}$ is given by
$$
P_{m,n}=\sum_{i=1}^{m}\sum_{j=1}^{n}E_{ij}\otimes E_{ij}^T,
$$
where $E_{ij}\in\mathbb{R}^{m\times n}$ has a one in the $(i,j)$ entry and every other entry is zero.
\end{enumerate}
\label{prop1}
\end{proposition}
\begin{proposition}~\cite{Astrom:70})
Let $x$, $y$ and $z$ be zero-mean random vectors with a jointly Gaussian distribution, and let $y$ and $z$ be independent. Also, let $S$ be
a symmetric matrix. Then the following facts hold:
\begin{enumerate}
\item[(a)] $\bold{E}\{x|y,z\}=\bold{E}\{x|y\}+\bold{E}\{x|z\}$\vspace{0.5mm}.
\item[(b)] $\bold{E}\{x|y\}=\bold{{Cov}}\{x,y\}\bold{{Cov}}^{-1}\{y,y\}y$\vspace{0.5mm}.
\item[(c)] $\bold{E}\{x^TSx\}=\bold{Tr}\left\{S\bold{{Cov}}\{x,x\}\right\}$.
\item[(d)] $\bold{E}\{x|y\}$ and $x-\bold{E}\{x|y\}$ are independent.
\end{enumerate}
\label{prop2}
\end{proposition}

\subsection{Proof Lemma 1}
To express the conditional estimate $\widehat{x}(k|k-1)$ in terms of  $\widehat{x}^{[1]}(k)$, we substitute Equation \eqref{200} into Equation \eqref{201} to eliminate $A\widehat{x}(k|k-1)+Bu(k)$. We have
\begin{align}
\widehat{x}(k|k-1)&=\widehat{x}^{[1]}(k)+\bigl(K(k-1)-{K}^{[1]}(k-1)\bigr)\bigl({y}(k-1)-C\widehat{x}(k-1|k-2)\bigr)\nonumber\\
&=\widehat{x}^{[1]}(k)+\bigtriangleup K(k-1)\widetilde{y}(k-1).\label{202}
\end{align}
Plugging $\widehat{x}(k|k-1)=x(k)-e(k)$ and $\widehat{x}^{[1]}(k)=x(k)-e^{[1]}(k)$ into Equation \eqref{202} leads to
\begin{align}
e^{[1]}(k)=e(k)+\bigtriangleup K(k-1)\widetilde{y}(k-1).\label{locale}
\end{align}
Since $e(k)$ is independent of $y(0:k-1)$, the two terms on the right hand side of Equation \eqref{locale} are independent. Thus,
\begin{align*}
P^{[1]}(k)&=\textbf{E}\left\{{e}^{[1]}(k){{e}^{[1]}(k)}^T\right\}\\
&=P(k)+\bigtriangleup K(k-1)\widetilde{Y}(k-1)\bigtriangleup K^T(k-1),\\
\widetilde{Y}^{[1]}(k)&=\textbf{E}\left\{\widetilde{y}^{[1]}(k){\widetilde{y}^{[1]}(k)}^T\right\}\\
&=CP^{[1]}(k)C^T+V,\\
\widetilde{P}(k)&=\textbf{E}\left\{{e}^{[1]}(k)\widetilde{y}^{\;T}(k-1)\right\}\nonumber\\
&=\bigtriangleup K(k-1)\widetilde{Y}(k-1).
\end{align*}

\subsection{Proof Lemma 2}
The independence between $x(k)-\widehat{x}(k)$ and $\widehat{x}(k)$ can be established by Proposition $\mbox{4}(\mbox{d})$. To calculate $\widetilde{x}(k)$, we proceed in three steps. First consider
\begin{align*}
u(k-1)=F(k-1) y(k-1)+G(k-1){y(k-2)}+f\bigl({y}(0:k-3)\bigr),
\end{align*}
where we used Equation \eqref{ucontrol}. Since $G(k-1){y(k-2)}+f\bigl({y}(0:k-3)\bigr)$ is a deterministic function of
$y(0:k-2)$, we have
\begin{align}
u(k-1)-\textbf{E}\left\{u(k-1)|y(0:k-2)\right\}&=F(k-1)\bigl(y(k-1)-\textbf{E}\{y(k-1)|y(0:k-2)\}\bigr)\nonumber\\
&=F(k-1)\widetilde{y}(k-1),\label{30}
\end{align}
where we used the definition of $\widetilde{y}$ (Equation \eqref{905}) to get the second equality. Second, consider 
\begin{eqnarray*}
\widehat{x}^{[1]}(k)&=A\widehat{x}(k-1|k-2)+B u(k-1)+{K}^{[1]}(k-1)\widetilde{y}(k-1),
\end{eqnarray*}
where we used Equation \eqref{200}. Since $\widehat{x}(k-1|k-2)$ is a linear function of $y(0:k-2)$, we have
\small
\begin{align}
\widehat{x}^{[1]}(k)-\textbf{E}\{\widehat{x}^{[1]}(k)|y(0:k-2)\}&={K}^{[1]}(k-1)\widetilde{y}(k-1)+B\bigl(u(k-1)-\textbf{E}\{u(k-1)|y(0:k-2)\}\bigr)\nonumber\\
&=({K}^{[1]}(k-1)+BF(k-1))\widetilde{y}(k-1),\label{31}
\end{align}
\normalsize
where we used the independence of $\widetilde{y}(k-1)$ and $y(0:k-2)$ to get the first equality, and Equation \eqref{30} to obtain the second equality. Finally, note that $x(k)={e}^{[1]}(k)+\widehat{x}^{[1]}(k)$. Thus,
\begin{align}
\widetilde{x}(k)&=x(k)-\textbf{E}\{x(k)|y(0:k-2)\}\nonumber\\
&={e}^{[1]}(k)+\bigl(\widehat{x}^{[1]}(k)-\textbf{E}\{\widehat{x}^{[1]}(k)|y(0:k-2)\}\bigr)\nonumber\\
&={e}^{[1]}(k)+\left({K}^{[1]}(k-1)+BF(k-1)\right)\widetilde{y}(k-1), \label{32}
\end{align}
where we used the independence of ${e}^{[1]}(k)$ and $y(0:k-2)$ to get the second equality and Equation \eqref{31} to obtain the last equality.

\subsection{Proof Lemma 3}
According to Proposition $\mbox{4}(\mbox{d})$, $\widehat{u}(k)$ is independent of $u(k)-\widehat{u}(k)$. Note that $v(k)$ is independent of
the previous outputs, so
\begin{align}
y(k)-\textbf{E}\{y(k)|y(0:k-2)\}&=v(k)+C\bigl(x(k)-\textbf{E}\{x(k)|y(0:k-2)\}\bigr)\nonumber\\
&=v(k)+C\bigl({e}^{[1]}(k)+\left(BF(k-1)+{K}^{[1]}(k-1)\right)\widetilde{y}(k-1)\bigr)\nonumber\\
&=\widetilde{y}^{[1]}(k)+C\bigl(BF(k-1)+{K}^{[1]}(k-1)\bigr)\widetilde{y}(k-1),\label{750}
\end{align}
where we used Equation (\ref{32}) to get the second equality and the definition of $\widetilde{y}^{[1]}$ (Equation\eqref{905}) to obtain the last equality. Since $f(y(0:k-2))$ is a linear function of $y(0:k-2)$, we have
\begin{align*}
\widetilde{u}(k)=&u(k)-\textbf{E}\{u(k)|y(0:k-2)\}\notag\\
=&F(k)\bigl(y(k)-\textbf{E}\{y(k)|y(0:k-2)\}\bigr)+G(k)\bigl(y(k-1)-\textbf{E}\{y(k-1)|y(0:k-2)\}\bigr)\nonumber\\
=&F(k)\left(\widetilde{y}^{[1]}(k)+C\bigl(BF(k-1)+{K}^{[1]}(k-1)\bigr)\widetilde{y}(k-1)\right)+G(k)\widetilde{y}(k-1)\nonumber\\
=&F(k)\widetilde{y}^{[1]}(k)+\bigl(F(k)C(BF(k-1)+{K}^{[1]}(k-1))+G(k)\bigr)\widetilde{y}(k-1),
\end{align*}
where we used Equation \eqref{750} and the definition of $\widetilde{y}$ (Equation \eqref{905}) to get the third equality. The proof is completed by defining
\begin{align*}
F^{[1]}(k)=G(k)+F(k)C\bigl({K}^{[1]}(k-1)+BF(k-1)\bigr).
\end{align*}

\subsection{Proof Lemma 4}
Due to the assumptions, $H(k)$ is positive definite and hence all terms in the $\widehat{J}$ are positive. Since $\widehat{u}(k)$ and $\widehat{x}(k)$ are functions of $y(0:k-2)$, the
optimal controller is given by (\ref{500}).

\subsection{Proof Lemma 5}
Let $A_j\in\mathbb{R}^{n\times m_j}$ denote the $j^{th}$ block column of matrix $A$. According to Proposition $\mbox{3(e)}$, we have
\begin{align*}
\textup{vec}(A_j)&=\textup{vec}\left(\begin{bmatrix} A_{1j}\\ \vdots \\ A_{pj}\end{bmatrix}\right)=P_{m_j,n}\textup{vec}\left(\begin{bmatrix} A^T_{1j} & \ldots & A^T_{pj}\end{bmatrix}\right)\\
&=P_{m_j,n}\begin{bmatrix} \textup{vec}(A^T_{1j}) \\ \vdots \\ \textup{vec}(A^T_{pj})\end{bmatrix}=P_{m_j,n}\begin{bmatrix} P_{n_1,m_j}\textup{vec}(A_{1j}) \\ \vdots \\ P_{n_p,m_j}\textup{vec}(A_{pj})\end{bmatrix}\\
&=P_{m_j,n}\textbf{diag}(P_{n_1,m_j},\ldots,P_{n_p,m_j})\begin{bmatrix} \textup{vec}(A_{1j}) \\ \vdots \\ \textup{vec}(A_{pj})\end{bmatrix}.
\end{align*}
Let $P_j=P_{m_j,n}\textbf{diag}(P_{n_1,m_j},\ldots,P_{n_p,m_j})$. Then
\begin{align}
\textup{vec}(A)=\begin{bmatrix} \textup{vec}(A_{1}) \\ \vdots \\ \textup{vec}(A_{q})\end{bmatrix}
=\underbrace{\textbf{diag}(P_1,\ldots,P_q)}_{P}\underbrace{\begin{bmatrix} \textup{vec}(A_{11}) \\ \vdots \\ \textup{vec}(A_{p1}) \\ \vdots \\ \textup{vec}(A_{1q}) \\ \vdots \\ \textup{vec}(A_{pq}) \end{bmatrix}}_{a_A}.\label{666}
\end{align}
Note that vector $a_A$ consists of all $pq$ sub-vectors $\textup{vec}(A_{11}),\ldots,\textup{vec}(A_{pq})$. Let $a^{\star}_A$ denote the vector containing only nonzero sub-vectors of $a_A$. We define $\mathcal{A}=\left\{i|[a_A]_i\neq 0\right\}$. Let $T_i=\begin{bmatrix} 0 & \ldots & I & \ldots & 0\end{bmatrix}^T$ be the block matrix with an identity in the $i^{th}$ block row. It is easy to see that there exists full column rank matrix $T$ whose columns are $T_j$ for
$j\in \mathcal{A}$ such that $a_A=Ta^{\star}_A$. This implies that Equation \eqref{666} can be written as
$$
\textup{vec}(A)=PTa^{\star}_A.
$$
The proof is completed by defining $E=PT$.

\subsection{Proof Lemma 6}
The equivalence of optimization problems follows simply by using the vec operator.
First note that $\text{vec}\bigl(F(k)\bigr)=E_1\xi_1(k)=\begin{bmatrix} I & 0\end{bmatrix}E\zeta(k+1)$. Using Proposition $\mbox{3(c)}$,
the first term on the right-hand side of \eqref{370} can be written as
\begin{align}
\textbf{Tr}\left\{H(k)F(k)VF(k)^T\right\}&= \text{vec}^T\bigl(F(k)\bigr)\bigl(V \otimes H(k)\bigr) \text{vec}\bigl(F(k)\bigr)\nonumber\\
&=  \zeta^T(k+1)E^T\begin{bmatrix}    I & 0\end{bmatrix}^T \bigl(V \otimes H(k)\bigr)\begin{bmatrix}    I & 0\end{bmatrix}E\zeta(k+1).\label{a1}
\end{align}
The second term on the right-hand side of \eqref{370} can be written as
\small
\begin{align}
&\textbf{Tr}\left\{H(k)\bigl(F(k)C-L(k)\bigr)P^{[1]}(k)\bigl(F(k)C-L(k)\bigr)^T\right\}= \text{vec}^T\bigl(F(k)\bigr)\bigl(CP^{[1]}(k)C^T \otimes H(k)\bigr) \text{vec}\bigl(F(k)\bigr)\nonumber\\
&\hspace{3.5cm}-2 \text{vec}^T\bigl(F(k)\bigr)\bigl(CP^{[1]}(k)\otimes H(k)\bigr) \text{vec}\bigl(L(k)\bigr)+ \text{vec}^T\bigl(L(k)\bigr)\bigl(P^{[1]}(k)\otimes H(k)\bigr) \text{vec}\bigl(L(k)\bigr)\nonumber\\
&\hspace{2cm}=  \zeta^T(k+1)E^T\begin{bmatrix}    I & 0\end{bmatrix}^T \bigl(CP^{[1]}(k)C^T \otimes H(k)\bigr)\begin{bmatrix}    I & 0\end{bmatrix}E\zeta(k+1)\nonumber\\
&\hspace{2.1cm}-2  \zeta^T(k+1)E^T\begin{bmatrix}    I & 0\end{bmatrix}^T \bigl(CP^{[1]}(k)\otimes H(k)\bigr)\text{vec}\bigl(L(k)\bigr)+ \textup{vec}^T\bigl(L(k)\bigr)\bigl(P^{[1]}(k)\otimes H(k)\bigr) \text{vec}\bigl(L(k)\bigr).\label{a2}
\end{align}
\normalsize

Likewise, $\textup{vec}\bigl(F^{[1]}(k)\bigr)=E_2\xi_2(k)=\begin{bmatrix} 0 & I\end{bmatrix}E\zeta(k)$ and
\small
\begin{align*}
\text{vec}\left(F^{[1]}(k)-L(k)\bigl({K}^{[1]}(k-1)+BF(k-1)\bigr)\right)&\nonumber\\
&\hspace{-3cm}=\text{vec}\bigl(F^{[1]}(k)\bigr)-\bigl(I \otimes L(k)B\bigr)\text{vec}\bigl(F(k-1)\bigr)
-\text{vec}\bigl(L(k){K}^{[1]}(k-1)\bigr)\nonumber\\
&\hspace{-3cm}=\begin{bmatrix} -I \otimes L(k)B & I \end{bmatrix}E\zeta(k)-\text{vec}\bigl(L(k){K}^{[1]}(k-1)\bigr),
\end{align*}
\normalsize
where we used Proposition $\mbox{3(a)}$ to obtain the second equality. The third term on the right-hand side of \eqref{370} can be written as
\small
\begin{align}
&\textbf{Tr}\left\{H(k)\left(F^{[1]}(k)-L(k)\bigl({K}^{[1]}(k-1)+BF(k-1)\bigr)\right)\vspace{-2mm}
\widetilde{Y}(k-1)\left(F^{[1]}(k)-L(k)\bigl({K}^{[1]}(k-1)+BF(k-1)\bigr)\right)^T\right\}\vspace{-2mm}\nonumber\\
&\hspace{2.8cm}=  \zeta^T(k)E^T\begin{bmatrix} -I \otimes L(k)B & I \end{bmatrix}^T\bigl(\widetilde{Y}(k-1)\otimes H(k)\bigr)\begin{bmatrix} -I \otimes L(k)B & I \end{bmatrix}E\zeta(k)\nonumber\\
&\hspace{3cm}-2  \zeta^T(k)E^T\begin{bmatrix} -I \otimes L(k)B & I \end{bmatrix}^T\bigl(\widetilde{Y}(k-1)\otimes H(k)\bigr) \textup{vec}\bigl(L(k){K}^{[1]}(k-1)\bigr)\nonumber\\
&\hspace{3cm}+ \textup{vec}^T\bigl(L(k){K}^{[1]}(k-1)\bigr)\bigl(\widetilde{Y}(k-1)\otimes H(k)\bigr) \text{vec}\bigl(L(k){K}^{[1]}(k-1)\bigr).\label{a3}
\end{align}
\normalsize
The last term on the right-hand side of \eqref{370} can be written as
\small
\begin{align}
&\hspace{-2cm}\textbf{Tr}\left\{H(k)\left(F(k)C-L(k)\right)\widetilde{P}(k)\left(F^{[1]}(k)-L(k)\bigl({K}^{[1]}(k-1)+BF(k-1)\bigr)\right)^T\right\}\nonumber\\
=&  \zeta^T(k)E^T\begin{bmatrix} -I \otimes L(k)B & I \end{bmatrix}^T \bigl(\widetilde{P}^T(k)C^T\otimes H(k)\bigr)
     \begin{bmatrix}    I & 0\end{bmatrix}E\zeta(k+1)\nonumber\\
&-  \zeta^T(k)E^T\begin{bmatrix} -I \otimes L(k)B & I \end{bmatrix}^T \bigl(\widetilde{P}^T(k)\otimes H(k)\bigr)
       \textup{vec}\bigl(L(k)\bigr)\nonumber\\
&-  \zeta^T(k+1)E^T  \begin{bmatrix}    I & 0\end{bmatrix}^T \bigl(C\widetilde{P}(k)\otimes H(k)\bigr)\textup{vec}\bigl(L(k){K}^{[1]}(k-1)\bigr)\nonumber\\
&+\textup{vec}^T\bigl(L(k){K}^{[1]}(k-1)\bigr)\bigl(\widetilde{P}^T(k)\otimes H(k)\bigr)\textup{vec}\bigl(L(k)\bigr).\label{a4}
\end{align}
\normalsize
Substituting \eqref{a1}-\eqref{a4} back into \eqref{370}, noting that $\widetilde{Y}^{[1]}(k)~=CP^{[1]}(k)C^T+V$, and omitting constant terms we arrive at \eqref{59}.

The proof can be completed by showing that $Z_1(k)$ is positive definite. Since $\widetilde{Y}^{[1]}(k)$, $\widetilde{Y}(k)$, and $H(k)$ are positive definite according to assumptions $\mbox{1}$ and $\mbox{2}$, $\widetilde{Y}^{[1]}(k-1)\otimes H(k-1)$ and $\widetilde{Y}(k-1)\otimes H(k)$ are positive definite according to Proposition $\mbox{3(b)}$. Therefore, Since $E$ has full column rank, $Z_1(k)$ is positive definite.

\subsection{Proof Lemma 7}
To prove the theorem, we start from the endpoint and iterate backwards in time. Define
\small
\begin{align}
\Pi(N)=\min_{\zeta(N)}\biggl\{{1 \over 2}\zeta^T(N)&{Z}_1(N)\zeta(N)
+\zeta^T(N-1){Z}_2(N-1)\zeta(N)-\zeta^T(N){b}(N)\biggr\}.\label{970}
\end{align}
\normalsize
Since $Z_1(N)$ is positive definite, by taking derivative with respect to $\zeta(N)$, the optimal value of $\zeta(N)$ is given by
\begin{align*}
\zeta^{\star}(N)=-R^{-1}(N)\bigl({Z}^T_2(N-1) \zeta(N-1)-c(N)\bigr),
\end{align*}
where $R(N)={Z}_1(N)$ and $c(N)={b}(N)$.

By substituting the optimal value of $\zeta(N)$ into Equation \eqref{970}, we have
\small
\begin{align*}
\Pi(N)=&-{1 \over 2}\zeta^T(N-1){Z}_2(N-1)R^{-1}(N){Z}^T _2(N-1)\zeta(N-1)\nonumber\\
&+\zeta^T(N-1){Z}_2(N-1)R^{-1}(N)c(N)-{1 \over 2}c^T(N)R^{-1}(N){c}(N).
\end{align*}
\normalsize
Note that the last term is constant and independent of $\zeta(N-1)$.

We proceed similarly and define
\small
\begin{align}
\Pi(N-1)=\min_{\zeta(N-1)}\biggl\{{1 \over 2}\zeta^T(N-1){Z}_1(N-1)\zeta(N-1)&+\zeta^T(N-2){Z}_2(N-2)\zeta(N-1)\nonumber\\
&\hspace{2.2cm}-\zeta^T(N-1){b}(N-1)+\Pi(N)\biggr\}\nonumber\\
&\hspace{-8cm}=\min_{\zeta(N-1)}\biggl\{{1 \over 2}\zeta^T(N-1)R(N-1)\zeta(N-1)+\zeta^T(N-2){Z}_2(N-2)\zeta(N-1)-\zeta^T(N-1)c(N-1)\biggr\}\label{971},
\end{align}
\normalsize
where 
\begin{align*}
R(N-1)&={Z}_1(N-1)-{Z}_2(N-1)R^{-1}(N){Z}^T_2(N-1),\\
c(N-1)&={b}(N-1)-{Z}_2(N-1)R^{-1}(N)c(N-1).
\end{align*} 
Equation \eqref{971} is the same as Equation \eqref{970}, but with the time arguments shifted one step. Thus,
$$
\zeta^{\star}(N-1)=-R^{-1}(N-1)\bigl({Z}^T_2(N-2) \zeta(N-2)-c(N-1)\bigr).
$$
The procedure can now be repeated, and
\begin{align*}
\Pi(1)=&\min_{\zeta(1)}\left\{{1 \over 2}\zeta^T(1)R(1)\zeta(1)-\zeta^T(1)c(1)\right\}.
\end{align*}
Therefore,
$$
\zeta^{\star}(1)=R^{-1}(1)c(1).
$$
\end{document}